\documentclass[%
 reprint,
 amsmath,amssymb,
 aps,
]{revtex4-2}

\usepackage{graphicx}
\usepackage{dcolumn}
\usepackage{bm}
\usepackage{comment}
\usepackage{caption}
\usepackage{subcaption}
\usepackage{url}

\usepackage{float} 

\usepackage{xurl}

\usepackage{xcolor}

\begin{document}

\preprint{APS/123-QED}

\title{Self-Similarity in Online Networks During Social Movements}

\author{Manuel Suarez-Roman}
 \affiliation{UNIE University, Madrid, Spain.}
 \affiliation{Valencian International University (VIU), Spain.}
 \affiliation{Universidad Carlos III de Madrid, Madrid, Spain.}
\author{M. \'Angeles Serrano}%
\affiliation{Departament de F\'isica de la Mat\`eria Condensada, Universitat de Barcelona, Mart\'i i Franqu\`es 1, E-08028 Barcelona, Spain}
\affiliation{Universitat de Barcelona Institute of Complex Systems (UBICS), Universitat de Barcelona, Barcelona, Spain}
\affiliation{ICREA, Passeig Llu\'is Companys 23, E-08010 Barcelona, Spain}

\author{Yérali Gandica}
\email{ygandica@gmail.com}
\affiliation{Facultad de Ciencias Económicas y Empresariales, Depto. Análisis Económico: Economía Cuantitativa.  C. Francisco Tomás y Valiente 5, 28049. \\ Autonomous University of Madrid, Madrid, Spain}
\date{\today}

\begin{abstract}
Online platforms provide an infrastructure for social movements, leaving digital traces that can be modelled as networks to quantify how information, participation, and coordination {develops during episodes of collective action}. In this work, we unveil {the scale-invariance of the connectivity structure of online social networks} through the analysis of three geographically and socio-politically distinct massive mobilisation events. By constructing co-occurrence networks from Twitter (now X) hashtag data and applying a degree-thresholding renormalisation procedure, we demonstrate that these highly correlated social phenomena exhibit clear signatures of {hyperbolic geometric organisation, which defines effective communication distances. At peak mobilisation times, this organisation exhibits enhanced self-similarity accompanied by more efficient navigability, mediated by a larger proportion of long range connections and more dominant hubs as compared with non-critical time windows. These critical-like points are also characterised by modular-to-nested transitions {in both the co-occurrence and bi-partite networks, as well as} maxima in user participation. These findings suggest the existence of common organising principles governing online networks during the peak mobilisation phases of social movements}.
\end{abstract}

\maketitle


\section{\label{sec:intro}Introduction} 

Cooperative social phenomena in online networks are short-lived bursts of coordinated behaviour in which many users align around a shared target or theme. This produces high correlations in activity. Users post and interact in synchronised bursts, reuse the same language, hashtags, memes, and URLs, and engage with the same accounts. This creates overlapping interaction patterns due to the shared goal of attracting attention and achieving a common objective within the short time span of an event. Large-scale studies have connected online social media behaviours to offline protest, showing that increased coordination of messages using {specific hashtags on social media, such as Twitter,} is associated with subsequent large-scale decentralised rotests~\cite{conover2013digital,alvarez2015sentiment,steinert2015online}. In network terms, structural signatures of these episodes include temporally clustered edges, denser and more cohesive communities, greater centralisation around hubs, and more similar user interaction profiles, with the coordination signal fading once the event window closes~\cite{borge2011structural,Esquirol2024,Beiro2024}. 

Intuitively, the structure of online social networks during peak mobilisation phases may suggest the emergence of critical-like states in which collective action becomes highly correlated across users and time \cite{Beiro2024}, so that small triggers can cascade into system-wide bursts of activity. However, these critical-like signatures remain largely uncharacterised.
In particular, scale invariance and other statistical scaling properties in multiscale network organisation during these mobilisation episodes remain largely unexplored at the topological level, and the relatively scarce existing work has mostly focused on dynamical characterisation and modelling~\cite{10.1145/2994142, 10.1162/978-0-262-31709-2-ch057}.

In complex networks, structural scale invariance has been investigated as self-similarity across length scales, adapting Mandelbrot's ideas to characterise certain geometric objects known as fractals~\cite{Mandelbrot1999}, which express the invariance of form under length rescaling. Such rescaling has been implemented in networks using a variety of definitions of internode distance~\cite{boguna2021network}. For instance, shortest path length was originally used in combination with box-counting coarse-graining to explore fractality and define fractal dimensions in networks~\cite{Song2005,song2006origins}. However, {using shortest-path length as a distance measure has serious limitations. Because complex networks contain long-range connections, shortest-path lengths are narrowly distributed around a well-defined mean that is typically very small relative to the number of nodes. For instance, an average distance of 3.57 has been reported among millions of Facebook users~\cite{edunov2016three}. Consequently, the small-world property restricts the range of available path-length scales, hindering the definition of concepts such as locality and distinguishability.} Hence, other approaches have been introduced. 

{Alternatively, Serrano et al. introduced a family of network models in hidden metric spaces, known as the $\mathbb{S}^D$ model, in which the probability that two nodes are connected decreases with their effective distance in an underlying hidden metric space~\cite{Serrano2008}. This latent geometry was subsequently shown to be effectively hyperbolic through the isomorphism between the $\mathbb{S}^D$ and $\mathbb{H}^{D+1}$ models~\cite{krioukov2010hyperbolic}, with the probability of connection depending on hyperbolic distances alone. Since then, a variety of network-embedding methods have been developed to produce maps of real networks in hyperbolic space~\cite{boguna2010sustaining,muscoloni2017machine,blasius2018efficient,garcia2019mercator,Jankowski2023}. This geometric approach has revealed self-similarity across scales as a fundamental symmetry of real networks, implying that the same connectivity law governs their structure at different scales.}

{One such transformation is geometric renormalisation~\cite{garcia2018multiscale}, which has shown that real networks exhibit self-similar structure across length scales~\cite{zheng2020geometric,zheng2021scaling,barjuan2025multiscale}. Another is a simple degree-thresholding renormalisation (DTR) scheme~\cite{Serrano2008}, which generates a nested hierarchy of subgraphs by progressively removing nodes whose degrees fall below a prescribed threshold (see \cite{gabrielli2025network} for a review of network renormalisation approaches). Many real-world networks display self-similarity along the DTR flow, once the nodes' degrees in the subgraphs are rescaled by the average degree of the corresponding subgraph. This symmetry, referred to here as multiscale $k$-cut self-similarity, together with high levels of clustering, provides a strong signature of an underlying spatial organisation as proven in~\cite{aliakbarisani2025clustering}}.

Motivated by the fact that online social networks are already characterised by high levels of clustering, due to the triadic closure mechanism that drives human interactions~\cite{Granovetter1977}, in this work we explored the {k-cut self-similarity} implied by latent geometry in complex networks. More precisely, we used the {DTR transformation} to search for signatures of multiscale $k$-cut self-similarity in the structural features of online social networks in the context of social movements. {We found evidence of self-similar structural patterns in the network organisation of online social movements across diverse geographical and socio-political contexts, consistent with a latent geometry. We therefore adopted the geometric modelling framework and used hyperbolic embeddings to obtain a geometric representation of the hashtag co-occurrence networks, thereby defining effective distances between hashtags. We further show that this organisation exhibits enhanced self-similarity accompanied by more efficient navigability at peak mobilisation times, mediated by a larger proportion of long range connections and more dominant hubs, as compared with non-critical time windows.}

The remainder of this paper is organised as follows. We continue with Section \ref{sec:data}, where we present the events for which the data were collected, as well as the methodology used to gather those data{, the network reconstruction process, and the DTR filtration method}. Then, in Section \ref{sec:results}, we report the empirical results and provide a comprehensive analysis of the observed outcomes obtained through the application of the methodology described in the section 
{\it Methods}, {included at the end of the manuscript}. Finally, in Section \ref{sec:conclusions}, we summarise the main findings and implications of our work.

\section{FROM DATA TO NETWORK ANALYSIS}
\label{sec:data}

\begin{table*}[t]
\caption{Description of the datasets employed.}
\label{tab:dataset}
\begin{tabular}{lrrrrr}

\hline
Dataset & \multicolumn{1}{c}{Timespan of Data Collection}  & \multicolumn{1}{c}{Seed hashtags} & \# hashtags & \# users & Country  \\ \hline
\textit{No al tarifazo} & 01/01/2019 6h – 07/01/2019, 23h & \#noaltarifazo and \#ruidazonacional & 136433 & 9670 & Argentina \\
\textit{9n} & 08/11/2019, 6h – 10/11/2019, 20h & \#9n and \#9ngranmarchaporlajusticia & 219488 & 8022 & Argentina\\
\textit{Charlie Hebdo} & 10/01/2015, 1h – 12/01/2015, 23h & \#charliehebdo and \#jesuischarlie & 185670 & 16146 & France \\
  
\hline
\end{tabular}
\end{table*}

{We analysed data collected from Twitter (now X) covering three
geographically and socio-politically distinct episodes of large-scale
mobilisation. In each case, online activity surged in the days surrounding the event and was accompanied by collective street action. Two
datasets correspond to Argentina. The first, \emph{No al tarifazo},
relates to a wave of nationwide protests in early January 2019 against
sharp increases in public utility and transport tariffs announced by the
government at the end of 2018; the demonstrations were organised through
social media and took the form of simultaneous \emph{cacerolazos}
(pot-banging protests) across the country \cite{lanacion2019ruidazo}. The second, \emph{9n}, refers to the mobilisation of 9 November 2019, in which citizens demanded that
the independence of the judiciary be preserved amid the impending change
of national government \cite{actualidadpolitica2019marcha9n}. The two Argentinian events differ markedly in nature: \emph{No al
tarifazo} was a socioeconomic protest, driven by grassroots
neighbourhood organisations and centred on the cost of essential
services, whereas \emph{9n} concerned an institutional matter: the
independence of the judiciary, and mobilised different sectors of society. 
The third dataset, \emph{Charlie Hebdo}, relates
to France in the aftermath of the terrorist attack on the offices of the
satirical magazine \emph{Charlie Hebdo} in Paris on 7 January 2015. In the days that followed, the attack generated a rapidly growing wave of online activity on Twitter, with hashtags such as $\#$CharlieHebdo and $\#$JeSuisCharlie, which culminated in a mass demonstration in defence of freedom of expression on 11 January 2015 \cite{wikipedia2015republicanmarches}. }

For each social movement, we identified the two most frequently used hashtags {(seed hashtags)} during the demonstration day(s) and compiled the set of users who posted at least one tweet containing either hashtag during that period. We then collected all tweets posted by these users over a broader observation window. Finally, we extracted from those tweets every hashtag appearing in their messages. {Tweets for the two Argentine mobilisations were restricted to those originating from Argentina, whereas tweets for the Charlie Hebdo dataset were restricted to those originating from France.} Table \ref{tab:dataset} summarises the seed hashtags used for data collection, and basic dataset statistics (number of users and number of distinct hashtags).

{Once all tweets were collected, we aggregated their associated hashtags into Temporal Windows (TWs), each consisting of all the hashtags posted between its start and end times. Because our analysis relies on a renormalisation procedure that progressively removes nodes from the co-occurrence network (see Methods), each TW must be large enough for the successive subgraphs to retain sufficient statistical mass. We therefore fixed the TW width, for each dataset, as the shortest width for which the co-occurrence network reaches at least 700 unique hashtags around the reference window in which the modular-to-nested transition was located in Ref.~\cite{Beiro2024}, and further characterised through entropic measures~\cite{Rico2024}. Because the datasets differ in density, this criterion yields a one-hour window for the \textit{No al tarifazo} mobilisation and two-hour windows for the \textit{9n} and \textit{Charlie Hebdo} movements.}

For each TW, we constructed weighted bipartite networks linking hashtags and users, where the weight of each edge represents the number of times a given user employed the corresponding hashtag during that temporal interval. We also built hashtag co-occurrence networks where hashtags are represented as nodes, and an undirected edge was added between any pair of hashtags co-used by at least one user within the same TW. Edge weights in the co-occurrence network were defined as the number of distinct users who posted both hashtags during that period, thus providing a user-based measure of co-occurrence strength. {Each network was filtered to remove hashtags tweeted by fewer than a certain number of users, thereby reducing noise whilst preserving the core structural features. Table~SI.1 presents the percentage of nodes and edges retained after this filtering process for each dataset and selected TW.} 

\begin{figure*}[!ht]
    \centering
    \includegraphics[width=\textwidth]{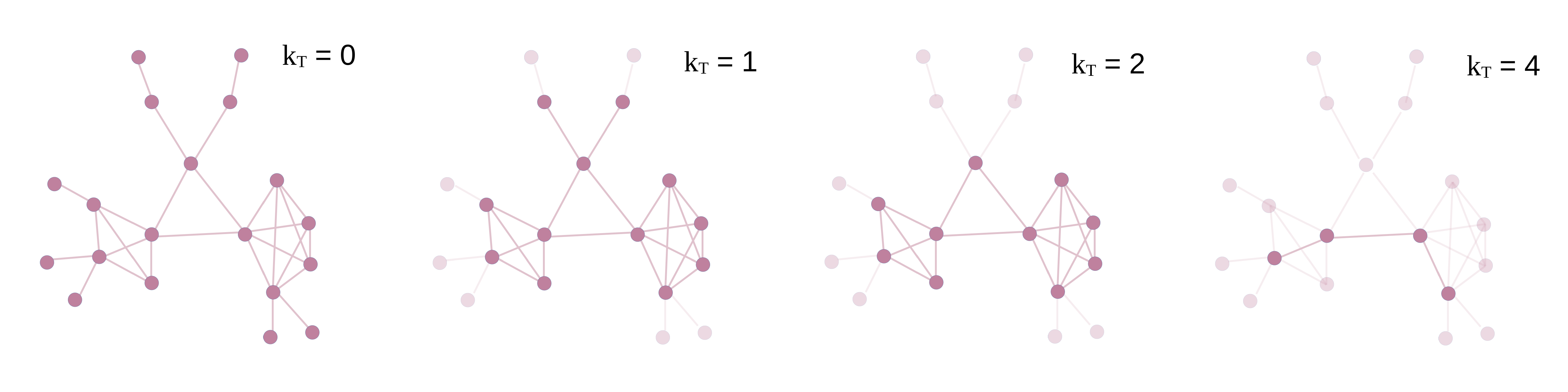}
    \caption{\textbf{Degree-thresholding renormalisation (DTR) procedure.} We can see the original network $(k_0=0)$ and three subsequent renormalised ones $k_1=1$, $k_2 = 2$ and $k_3=4$, where the nodes with {$k_T$} links or less were removed from the network at the $T$-th iteration.}
    \label{fig:renormalization}
\end{figure*}

{We applied degree-thresholding renormalisation (DTR) to the filtered networks. The procedure was introduced in~\cite{Serrano2008} to explore the multiscale organisation of networks, and is applied here to the hashtag co-occurrence networks in each TW. This method compares the connectivity structure of the original network with that of the subgraphs obtained at different scales}, once the degrees of the nodes in each subgraph are rescaled by the average degree of that subgraph. More specifically, given a graph $G$, DTR produces a nested hierarchy of $n$ subgraphs {$G(k_T)$} $\subseteq G$ obtained by removing all nodes from $G$ whose degree is equal to or less than {$k_T$}, where {$k_T = 0, 1, 2, \ldots, n$} denotes the sequence of degree thresholds effectively filtering out less connected nodes. Figure~\ref{fig:renormalization} illustrates this procedure with a toy example.

\label{sec:dtr}

\section{\label{sec:results}Results}

\subsection{Detecting the Critical Temporal Window}
\label{sec:resultdetecting}

\begin{figure*}[t]
    \centering
    \includegraphics[width=0.98\linewidth]{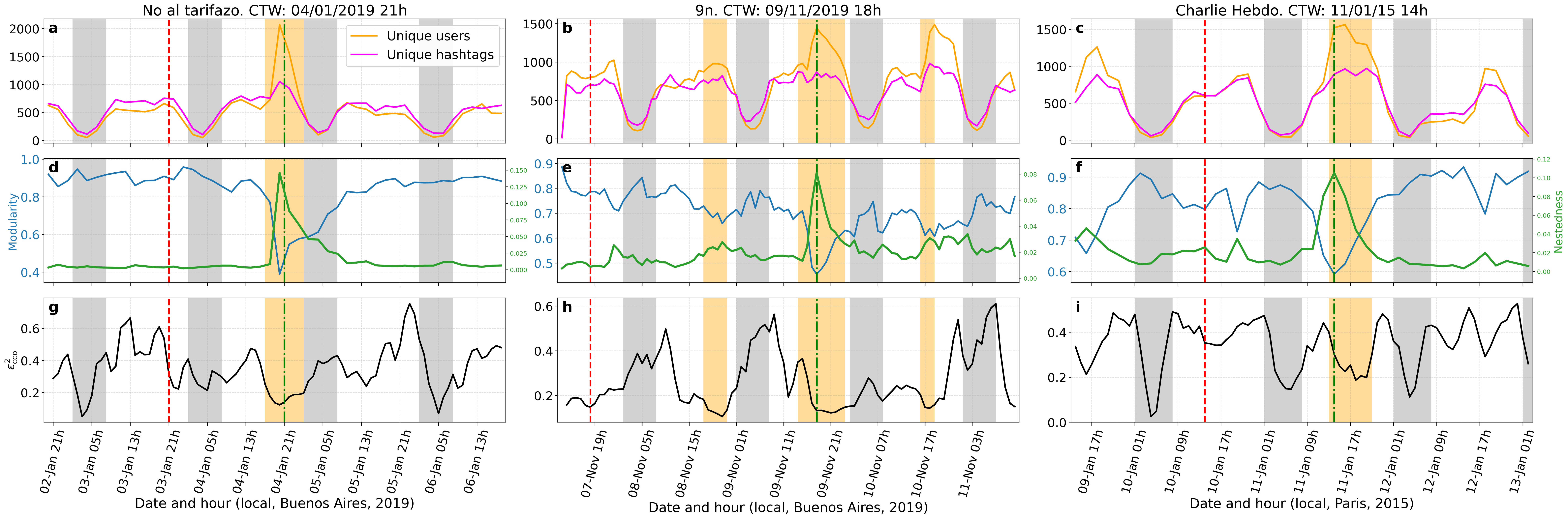}
    \caption{\textbf{Detecting the Critical Temporal Window (CTW).} {(a-c) Number of unique users (yellow) and unique hashtags (pink) across the temporal windows for the three social movements. Grey bands indicate low-activity periods (01:00--08:00 local time) excluded from analysis. (d-f) Evolution of modularity (blue) and nestedness (green) in bipartite user--hashtag networks. (g-i) Temporal evolution of the $\epsilon^2_{cco}(k)$ self-similarity metric, smoothed using a moving average over $N=3$ consecutive windows.} Yellow bands in all panels highlight intervals with minimum $\epsilon^2_{cco}(k)$ values, defining candidates for the Critical Temporal Windows. The identified CTW for each movement is marked by a {green} segmented line. For \textit{No al Tarifazo} (left) and \textit{Charlie Hebdo} (right), a single clear CTW emerges. For \textit{9n} (centre), three candidate segments show low $\epsilon^2_{cco}(k)$ values, but only the second exhibits a clear transition in the bipartite networks, unambiguously identifying the CTW. {The red dashed lines mark non-critical windows used in the analysis carried out in the SI. All times are given in the local time of the corresponding country (Buenos Aires, GMT$-$03:00, for \emph{No al Tarifazo} and \emph{9n}; Paris, GMT$+$01:00, for \emph{Charlie Hebdo})}.
    }
  \label{fig:detecting}
\end{figure*}

{To detect the Critical Temporal Window for each mobilisation, we first examine the temporal evolution of activity. Figures~\ref{fig:detecting}a-c show the number of unique hashtags and unique users across the TWs for the three events. We first exclude low-activity periods from all subsequent analyses, as these windows contain sparse participation and limited relevant content that could bias the identification of the CTW. These periods correspond to the intervals between \texttt{01:00} and \texttt{08:00} local time in the respective countries where each social movement took place, and are marked in grey in Figure~\ref{fig:detecting}. Within the active periods, there is a single clear dominant peak of engagement in \textit{No al tarifazo}, whilst \textit{Charlie Hebdo} and \textit{9n} display several maxima associated with moments of maximum online activity.}

{For each mobilisation, we computed the evolution of activity, modularity, and nestedness across all TWs during the period of analysis. The unique-hashtag and unique-user counts (Figs. ~\ref{fig:detecting}a–c) are consistent with those reported in Refs. ~\cite{Beiro2024,Rico2024} for the same three events. For modularity and nestedness (Figs. ~\ref{fig:detecting}d–f), however, we use the bipartite user–hashtag network, whilst the references mentioned used the unipartite hashtag co-occurrence network; the corresponding unipartite analysis is provided for comparison in Fig. SI.1.}

{To quantify the similarity between the original and DTR filtered networks, we used the $\epsilon^2_{cco}$ collapse-quality metric introduced in ~\cite{aliakbarisani2025clustering} (see Methods~C for a detailed explanation). The $\epsilon^2_{cco}$ is computed by averaging the squared relative differences between the clustering spectrum of the original network and those of the corresponding subgraphs over the range of rescaled degrees shared by both. An analogous collapse metric for the complementary cumulative degree distribution function is not adopted here, because the degree distribution already collapses under degree-thresholding renormalisation in scale-free configuration-model networks~\cite{Serrano2008}.}

{Figs.~\ref{fig:detecting}g-i show the temporal evolution of the $\epsilon^2_{cco}(k)$ self-similarity metric (lower for better collapse of the rescaled curves) for each movement, smoothed using a moving average over $N = 3$ consecutive TWs to reduce noise. As shown in Fig.~SI.2, the raw series fluctuate considerably, and the moving average attenuates this high-frequency noise to allow robust identification of the global minima.}

The minima in $\epsilon^2_{cco}(k)$ {help to identify} candidate CTWs. For \textit{No al Tarifazo} and \textit{Charlie Hebdo}, a single clear minimum emerges, which coincides precisely with engagement peaks (Fig.~~\ref{fig:detecting}g,i) and well-defined modular-to-nested transitions in the bipartite networks (Figs.~~\ref{fig:detecting}d,f), unambiguously defining unique candidate CTWs highlighted in yellow across all panels.

For \textit{9n}, however, three segments show low $\epsilon^2_{cco}(k)$ values (Fig.~~\ref{fig:detecting}h). Notice that unipartite co-occurrence networks (Fig. SI.1) do not help to resolve this ambiguity, as they also exhibit two candidate windows with modular-to-nested transitions, making it difficult to unambiguously identify the CTW. The bipartite networks (Fig.~\ref{fig:detecting}h), however, resolve this ambiguity: only the second segment exhibits both a sharp drop in $\epsilon^2_{cco}(k)$ and a clear modular-to-nested transition, unambiguously identifying Saturday, November~9, 2019 from 18:00 to 20:00 (Buenos Aires time, GMT$-03{:}00$) as the CTW (marked with a dashed line in all corresponding panels of Fig.~\ref{fig:detecting}). 

For the \textit{No al tarifazo} movement, the transition spans two consecutive TWs in the co-occurrence representation; however, the bipartite networks resolve it into a single, well-defined peak, allowing us to identify the CTW as Friday, January~4, 2019 from 21:00 to 22:00 (Buenos Aires time, GMT$-03{:}00$). Finally, using the same reasoning, for the \textit{Charlie Hebdo} dataset, we set the CTW to Sunday, January~11, 2015 from 14:00 to 16:00 (Paris time, GMT$+01{:}00$). The bipartite representation thus provides a clearer signal for detecting the CTWs and is used here as the primary guide for discerning the transitions.

{For comparison, we also selected non-critical time windows (NCTWs), marked by red dashed lines in Fig.~\ref{fig:detecting}. Note that the threshold used for edge-weight filtering differs between the CTW and NCTW networks. This difference arises because the original CTW and NCTW networks have substantially different numbers of nodes and edges. Consequently, for each NCTW network we selected the edge-weight threshold that yielded a filtered network closest in size to its CTW counterpart. This ensures that the resulting networks remain comparable to their CTW counterparts in absolute node and edge counts. The specific values are given in Table SI.1, which also shows that the maximum degree is substantially larger in the CTW networks than in the NCTWs. }

\begin{figure*}[t] 

  \centering
  \includegraphics[width=\textwidth]{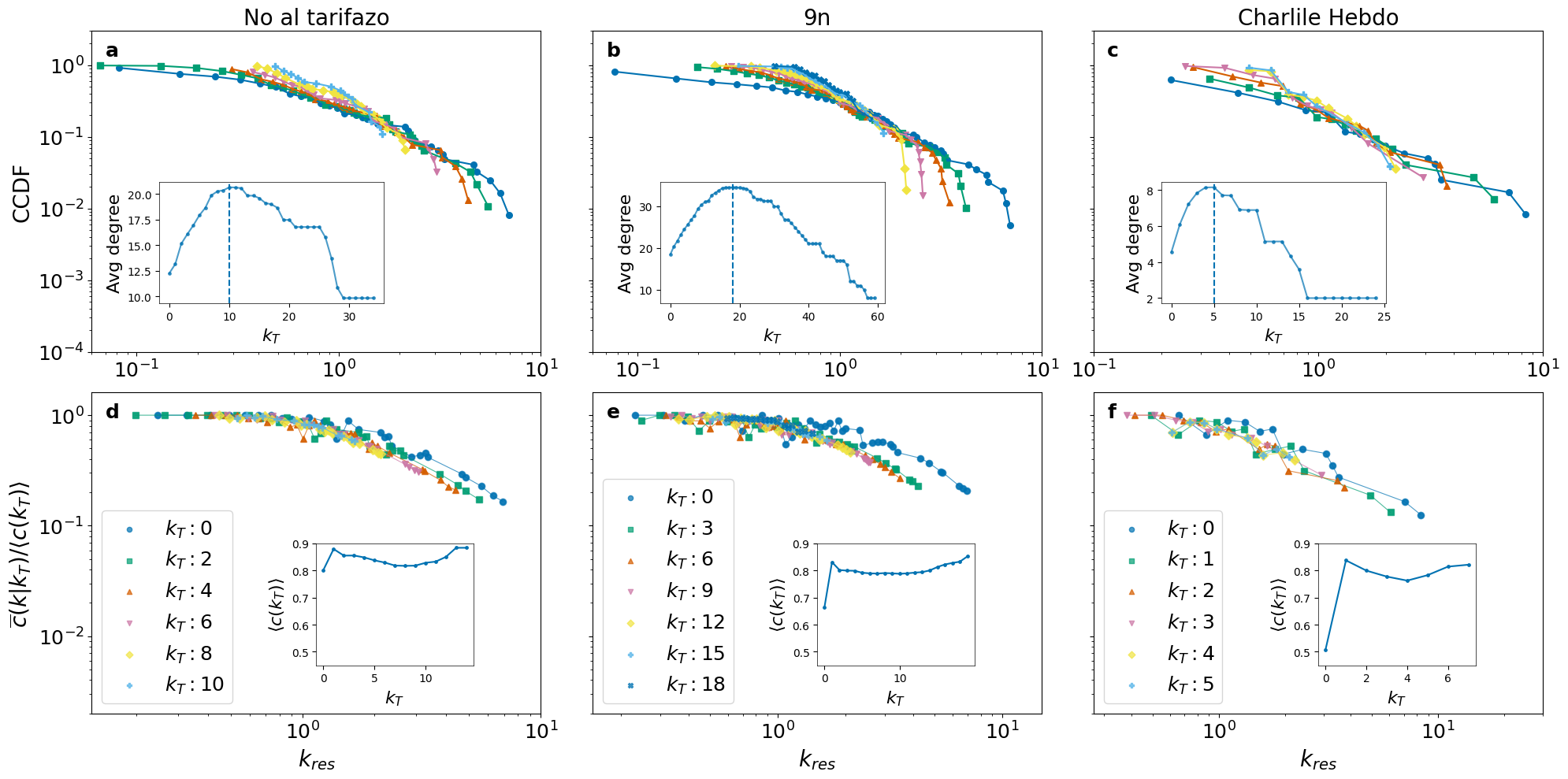}
\caption{\textbf{Complementary cumulative degree distribution and average clustering coefficient in the CTW}.The top row presents the complementary cumulative degree distribution function (CCDF) as a function of the rescaled degree, $k_{res}$, for various values of the degree threshold, {$k_T$}. The inset illustrates how the average degree of the resulting networks increases with {$k_T$}.A blue dashed line indicates the cut-off point beyond which the results are excluded from further analysis due to finite-size effects forcing the average degree to decrease. The bottom row displays the spectrum of the average clustering coefficient rescaled by the average clustering coefficient of the corresponding subgraph as a function of the rescaled degree. 
}
    \label{fig:self_similarity}
\end{figure*}

\subsection{Self-similarity across the DTR sequence}

{We calculated two main topological properties of each subgraph in the DTR flow of the networks during the identified CTWs and selected NCTWs: the degree distribution and the clustering spectrum.} In Figs.~\ref{fig:self_similarity}a-c, we present the complementary cumulative degree distribution function (CCDF), which quantifies the probability that a node has degree no smaller than a certain value, for a selection of DTR subgraphs of each dataset in its CTW. For each curve, the x-axis shows the degree rescaled by the average degree of the corresponding subgraph, which we call $k_{res}$. The insets show how the average degree of the resulting subgraphs increases with the applied threshold $k_T$ until it reaches a maximum (marked by a vertical dashed line), after which it decreases due to finite-size effects. We therefore stop the sequence of subgraphs at this maximum. {The corresponding cut-off thresholds are $k_T=10$, $18$, and $5$ for No al tarifazo, 9n, and Charlie Hebdo, respectively.} Although the curves span a limited range of values, their collapse is clearly evident. {As explained in Section~\ref{sec:resultdetecting}, we do not report an analogous $\epsilon^2$ metric for these distributions.}

We also calculated the clustering spectrum $\bar{c}(k)$ by averaging the local clustering coefficients $c_i=2T_i/(k(k-1))$ of nodes in the same degree class $k$, where $T_i$ is the number of triangles in which a node $i$ participates. In Figs.~\ref{fig:self_similarity}d-f, we show the clustering spectrum of the subgraphs as a function of the degree rescaled by the average degree of the corresponding subgraph. Remarkably, we observe a clear collapse of the clustering coefficient onto a single functional form across renormalisation steps for all three social movements, {consistent with the $\epsilon^2_{cco}$ minima identified at the CTW in Figs.~\ref{fig:detecting}g-i. This collapse provides evidence of a robust self-similar behaviour in the network structure over multiple scales, where local patterns of organisation are replicated at progressively larger scales.}

The slight variations in the range of {$k_T$} values reflect differences in network size and density among the three events, yet the fundamental self-similar behaviour remains consistent. Note that the original curve deviates slightly, indicating distinct statistical behaviour for the lowest-degree nodes.

{The observed approximate scale-invariant behaviour suggests that the underlying organisational principles governing hashtag co-occurrence during critical mobilisation periods persist across different levels of degree thresholding and are thus largely independent of degree scale. Crucially, this property is markedly stronger at the CTW. In the NCTWs, the CCDF and clustering spectrum curves collapse less clearly across thresholds (Fig.~SI.3, right column), and $\epsilon^2_{cco}$ is substantially higher than at the corresponding CTW within each dataset (0.4591 vs.\ 0.1549 for No al tarifazo, 0.2996 vs.\ 0.134 for 9n, and 0.4557 vs.\ 0.3025 for Charlie Hebdo). This suggests that enhanced multiscale self-similarity is a distinctive structural signature of the CTW.}

\subsection{Network embedding in hyperbolic space}

\begin{figure*}[t]
    \centering
    \begin{subfigure}[b]{0.32\textwidth}
     \caption{\centering No al tarifazo}
        \includegraphics[width=\linewidth]{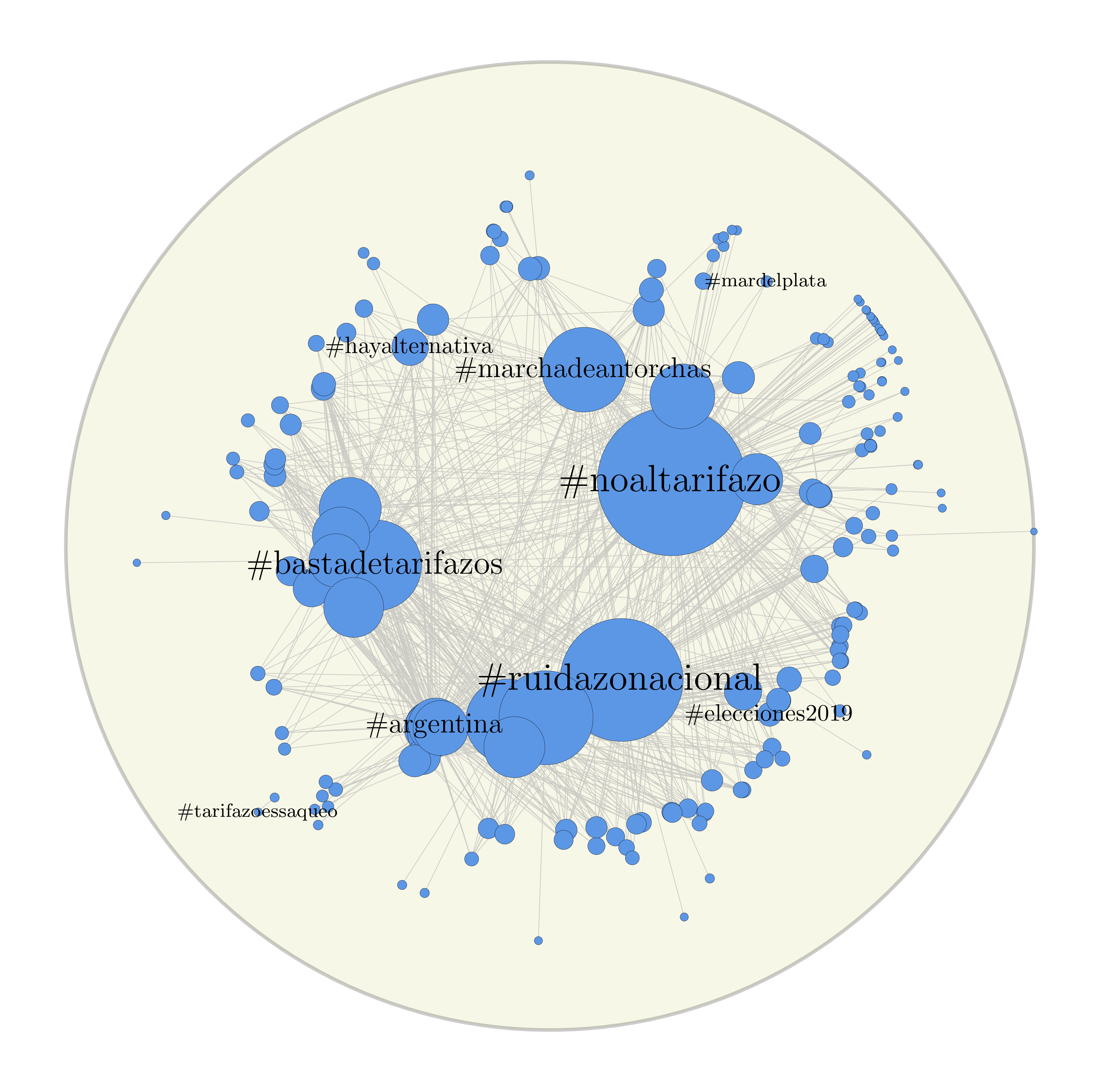}
    \end{subfigure}
    \hfill
    \begin{subfigure}[b]{0.32\textwidth}
      \caption{\centering 9n}
        \includegraphics[width=\linewidth]{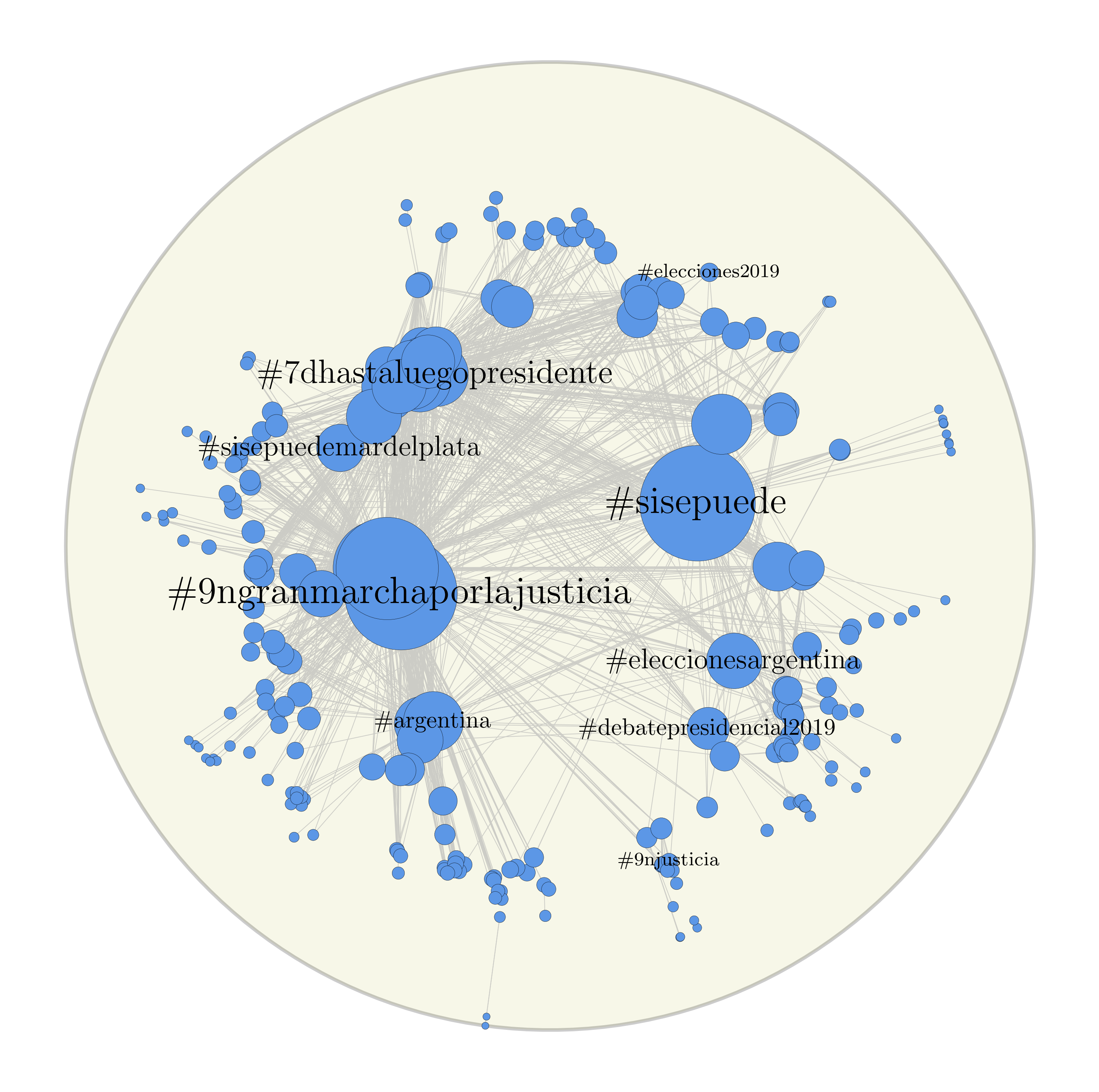}
    \end{subfigure}
    \hfill
    \begin{subfigure}[b]{0.32\textwidth}
      \caption{\centering Charlie Hebdo}
        \includegraphics[width=\linewidth]{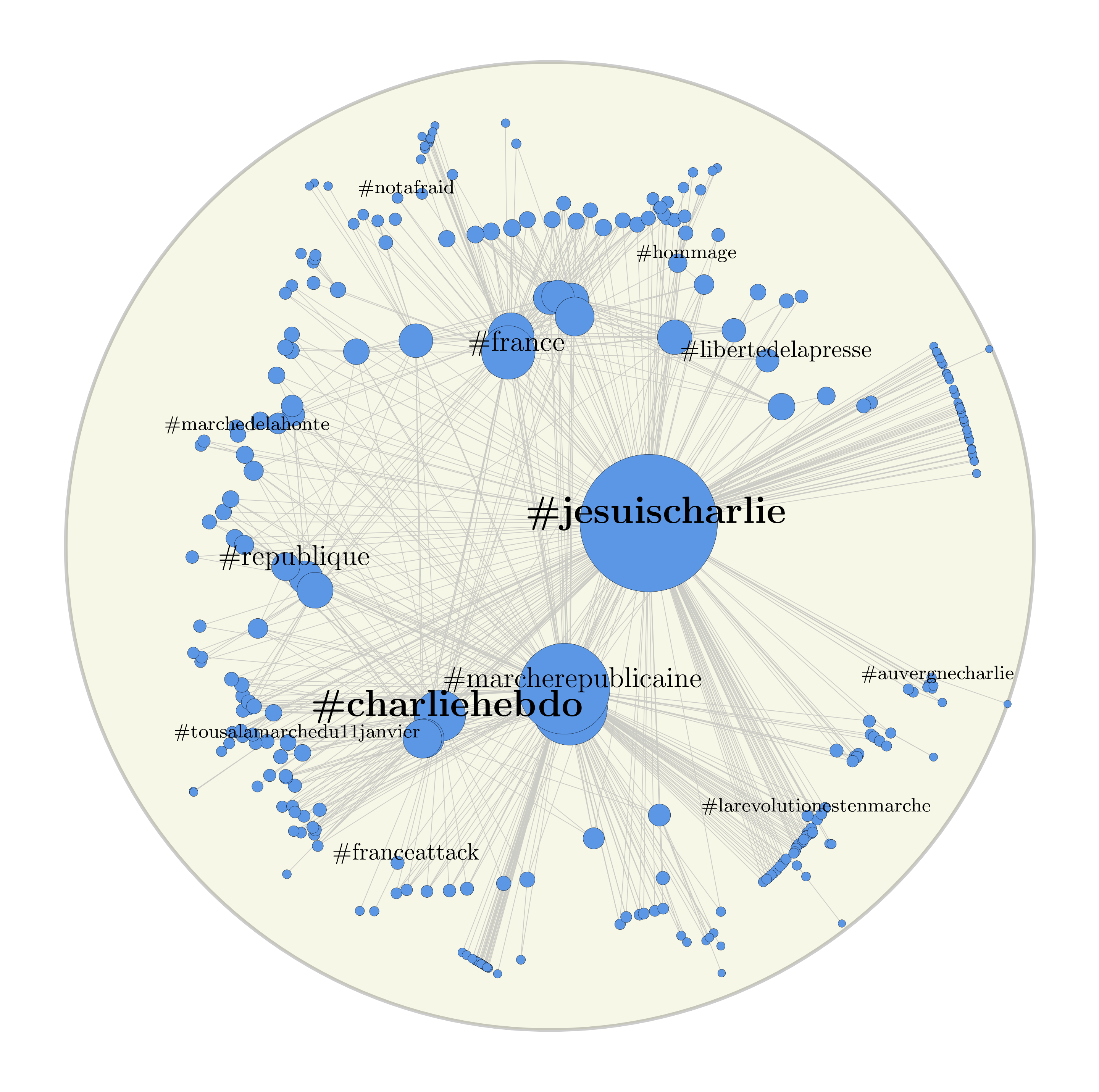}
    \end{subfigure}
    \vspace{-0.25cm} 

    \begin{subfigure}[b]{0.32\textwidth}
        \includegraphics[width=\linewidth, height=4cm]{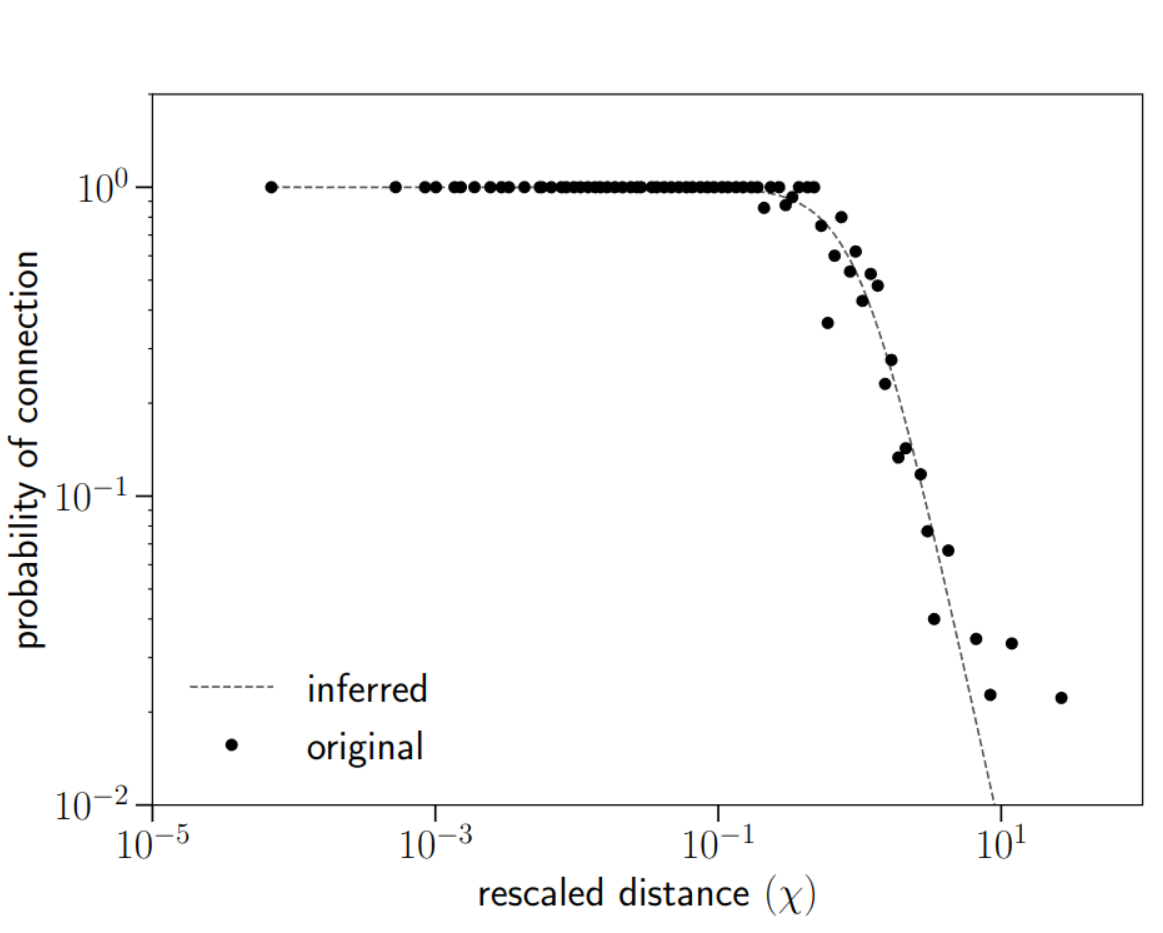}
        \end{subfigure}
    \hfill
    \begin{subfigure}[b]{0.32\textwidth}
        \includegraphics[width=\linewidth, height=4cm]{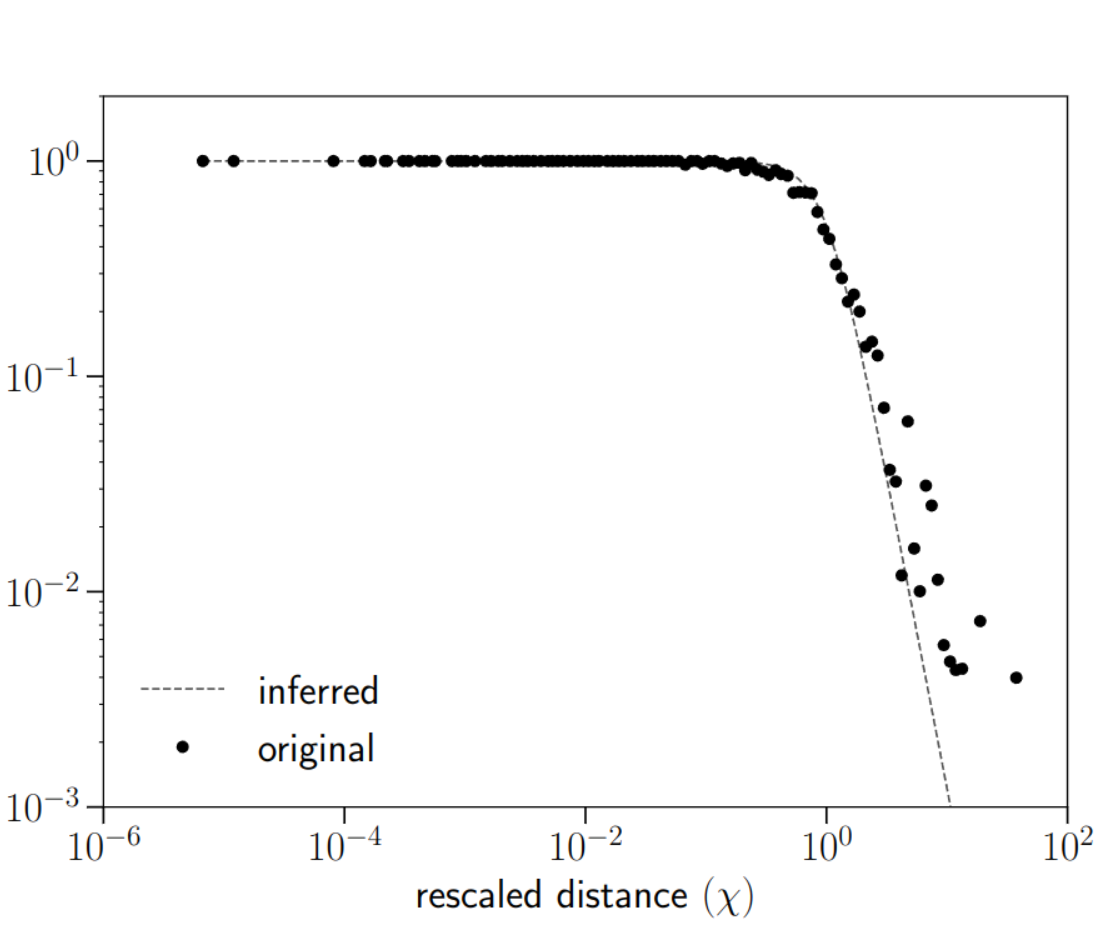}
        \end{subfigure}
    \hfill
    \begin{subfigure}[b]{0.32\textwidth}
        \includegraphics[width=\linewidth, height=4cm]{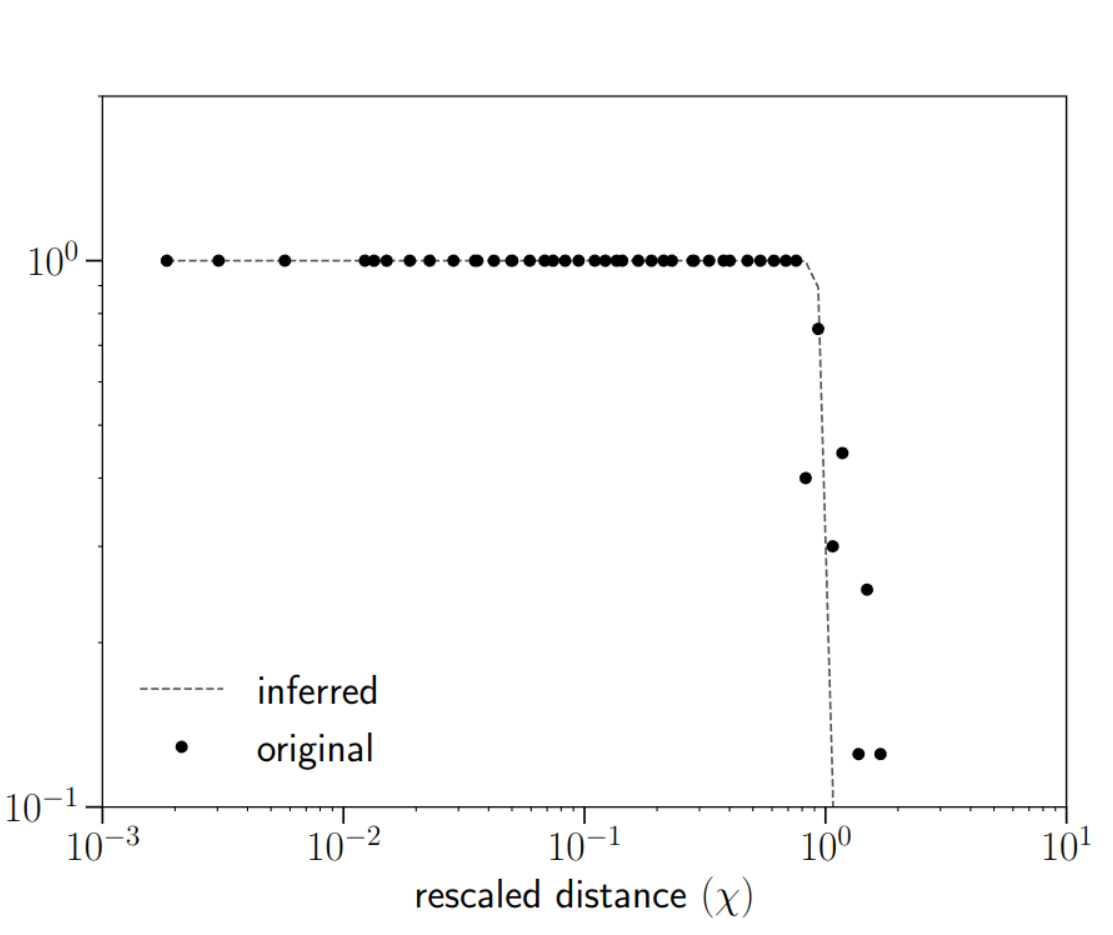}
        \end{subfigure}
    \caption{\textbf{Hyperbolic embeddings {of the hashtag co-occurrence networks} and connection probability profiles.} The top row presents the two-dimensional projections of the hyperbolic embeddings obtained using the D-Mercator method for the filtered Critical Temporal Window networks of the three social movements. Node size is proportional to degree, and spatial proximity reflects the likelihood of connection in the underlying hyperbolic geometry. The bottom row displays the connection probability as a function of rescaled hyperbolic distance for each dataset. The dotted line represents the inferred model, whilst black points correspond to the original network data. The excellent agreement between the model and empirical data demonstrates that the co-occurrence networks possess latent metric structures consistent with hyperbolic geometry, where connection probability decays smoothly with hyperbolic distance. From left to right: No al tarifazo, 9n, and Charlie Hebdo datasets. Prominent hashtags such as \texttt{\#noaltarifazo}, \texttt{\#9ngranmarchaporlajusticia},
and \texttt{\#jesuischarlie} are visible near the centre of each embedding, consistent with their high degree and central role in the mobilisation.}
    \label{fig:dmercator}
\end{figure*}

{The high level of clustering, together with the approximate multiscale $k$-cut self-similarity observed at the CTWs and, to a lesser extent, at the NCTWs, points to connectivity constrained by geometry as a potential explanation for the observed symmetry ~\cite{aliakbarisani2025clustering}.} Here we use the geometric $\mathbb{S}^1$ network model, which generates networks with clustering and multiscale $k$-cut self-similarity, {in which} nodes have a popularity coordinate directly controlling observed degrees and a similarity coordinate implemented as an angular position in a one-dimensional sphere or circle. Connections are formed with likelihoods that increase with popularity and decrease with angular distance. 

{Taking the network topology as input, we employed the D-Mercator tool~\cite{Jankowski2023} to embed the CTW and NCTW} {networks into their latent hyperbolic geometry. D-Mercator calculates the popularity and similarity coordinates of the nodes that give the maximum congruence between the $\mathbb{S}^1/\mathbb{H}^2$ model and the observed topology, and also provides estimates of the radius of the hyperbolic map and the model parameters $\beta$ and $\mu$ controlling, respectively, the level of clustering and the average degree in synthetic networks produced by the model.}

Figure \ref{fig:dmercator} displays the hyperbolic embeddings of the CTW networks for the three social movements, alongside the comparison of their empirical and inferred connection probability curves. {The corresponding validation plots and the inferred values of $\beta$ and $\mu$ are provided in Fig. SI.6 and Table SI.2, respectively, of the Supplementary Information. The successful embeddings provide strong evidence that our co-occurrence networks can be explained by a latent geometric structure satisfying the triangle inequality.} The upper row shows the two-dimensional hyperbolic embeddings, where node size represents degree and spatial proximity in the hyperbolic plane reflects the likelihood of connection. In each embedding, the seed hashtags---such as \texttt{\#noaltarifazo} and \texttt{\#ruidazonacional}, \texttt{\#9ngranmarchaporlajusticia}, and \texttt{\#jesuischarlie} and \texttt{\#charliehebdo}---occupy central positions with the lowest radial coordinates, reflecting their high popularity and role as focal points of collective attention. Surrounding them, thematically related hashtags cluster at small angular distances, indicating high similarity --- for instance, \#sepuede in the \textit{No al tarifazo} embedding, and \#marcherepublicaine and \#libert\'e\'egalit\'efraternit\'e in the \textit{Charlie Hebdo} embedding --- whilst more peripheral or loosely related hashtags are pushed towards the outer regions of the hyperbolic disc. The lower row presents the connection probability as a function of rescaled hyperbolic distance, comparing the inferred model (dotted line) against the empirical network data (black points). For all three datasets, we observe excellent agreement between the inferred hyperbolic geometry and the original network structure. The connection probability decays smoothly with hyperbolic distance, closely following the expected behaviour for networks embedded in underlying metric spaces. 

{The \(S^1\) geometric model also reproduces several topological observables well in NCTWs, indicating that hyperbolic representability is not exclusive to the CTW. Validation plots for the selected NCTWs are in Fig. SI.7. In the NCTWs, by contrast, the values of $\beta$ are consistently larger than in the CTWs (Table SI.2), indicating a reduced proportion of long-range connections in favour of higher local connectivity. However, multiscale self-similarity is markedly stronger at the CTWs in all three datasets: the \(\epsilon^2_{cco}\) values are substantially lower and the degree and clustering curves collapse more clearly across scales.}

Moreover, {on the hyperbolic maps, greedy routing exhibits higher navigability during periods of critical mobilisation, with success ratios between 0.950 and 0.983 and {average geometric stretch} between 1.346 and 1.543 (Table SI.2). Regardless of the specific context, whether protests against government policies in Argentina or solidarity marches following a terrorist attack in France, the underlying network structures, generated by the joint use of hashtags, thus support close-to-optimal routing between users, consistent with a self-organisation process that favours efficient message diffusion.}

{Thus, the distinctive structural signature of CTWs is not the mere existence of a hyperbolic embedding, but the enhancement of scale-invariant organisation, accompanied by a more efficient basis for greedy navigation.} 

\section{\label{sec:conclusions} Discussion}
In this work, we have {investigated self-similar} structural patterns in social movements through the analysis of three distinct mobilisation events: the 9n justice reform protests and No al tarifazo demonstrations in Argentina, and the Charlie Hebdo marches in France. By applying the DTR degree-thresholding renormalisation procedure to co-occurrence networks constructed from {Twitter (now X)} data, we have demonstrated that these highly correlated social phenomena exhibit clear signatures of self-similarity{, especially} at critical periods. The observation that self-similarity {is present} across all three datasets, despite their differing geographical and socio-political contexts, points towards {common organising principles governing online networks during social movements.}

Our methodology successfully identified Critical Temporal Windows (CTWs) for each movement. The convergence of multiple indicators—including the maxima in user and hashtag participation, the modular-to-nested transition (in both bipartite and co-occurrence networks), {the $\epsilon^2_{cco}$ collapse-quality metric indicating enhanced self-similarity, and accompanied by more efficient navigability in hyperbolic embeddings mediated by a larger proportion of long-range connections and more dominant hubs than in NCTWs}—provides robust evidence that these CTWs represent distinct {regimes} in the dynamics of social mobilisation. Notably, the identified CTWs coincide {with the physical demonstrations already taking place and, more precisely, with the moments of peak online activity}, suggesting that the underlying network structures reflect fundamental organisational principles of collective attention. 

{In this context, ``organisational principles of collective attention" refers to the self-organisation of attention within the network structure, that is, the modular-to-nested reorganisation and the {enhanced self-similarity that develops} when attention concentrates. It does not refer to the logistical organisation of the demonstrations. We make no causal or predictive claim regarding the physical events. The alignment we report is a temporal co-occurrence between the window in which the network adopts a structure supporting efficient communication and the timing of the physical demonstrations, and not an assertion that the network signature drives or precedes it.}

The highly clustered nature of social interactions, driven by triadic closure mechanisms, is reflected in high levels of clustering. The clustering coefficient curves, as well as the degree distributions, collapse onto single curves obtained for the subgraphs generated by the DTR renormalisation procedure. {The simultaneous occurrence of high levels of clustering and DTR self-similarity at the CTWs reduces the family of models that can be used to explain these observations. On the other hand, one would like the models to be maximum entropy models, such that networks in the corresponding ensemble are maximally random except for the assumed constraints. As far as we know, the $\mathbb{S}^{D}/\mathbb{H}^{D+1}$ is the only family of models that produce networks with high levels of clustering and DTR self-similarity whilst being maximally random ensembles, as proved in~\cite{aliakbarisani2025clustering} and~\cite{angeles2026statistical}. Hence, our findings suggest that these networks possess a latent hyperbolic metric structure}. The successful embeddings using the D-Mercator method reinforce the notion that in these systems the likelihood of interaction decreases with an effective distance, which combines similarity and popularity dimensions. { These embeddings are also highly navigable under greedy routing, {especially at peak mobilisation times} as shown in Section~III.C, consistent with a self-organisation process that favours efficient {greedy routing navigability}. Our interpretation is then that this scale-invariant behaviour, and its consistency with maximally random $\mathbb{S}^D/\mathbb{H}^{D+1}$ ensembles, emerges from the co-ordinated efforts of participants to achieve common goals within compressed time frames.}

These results have important implications for understanding the organisational dynamics of social movements. The presence of self-similarity indicates that local patterns of coordination and information diffusion are replicated at progressively larger scales.
This hierarchical self-organisation may represent an optimal structural configuration for balancing information dissemination and network coherence during critical periods. Our results suggest that analysing other social phenomena characterised by peak correlation in attention and collective actions would be valuable for advancing our understanding of the fundamental principles governing criticality in social systems.

\section*{\label{sec:methodology}Methods}
\setcounter{figure}{0} 
\renewcommand{\thefigure}{SI.\arabic{figure}} 

{\subsection{Data Gathering}
Twitter/X data for the events analysed in this study were collected as part of a separate published study~\cite{Beiro2024}. Data for the Argentine events (No al Tarifazo and 9N) were obtained by a collaborator through the Twitter/X Academic Research API. Data for the Charlie Hebdo event were acquired commercially. The No al Tarifazo dataset is publicly available at ~\cite{Gandica2026}. The 9N and Charlie Hebdo datasets are not publicly available.}

\subsection{Modular-to-nested transitions}
\label{sec:mod_nest_calc}

{We characterise the modular-to-nested transition through two
complementary quantities: the modularity of the network's community
partition and its nestedness. Modularity measures how well a partition
divides the network into internally dense, mutually sparse communities.
For a graph $G = (V, E)$, with $V$ the set of nodes and $E$ the set of
edges, it is defined as~\cite{Newman2010}
\begin{equation}
  Q = \frac{1}{2m} \sum_{ij}
      \left( A_{ij} - \frac{k_i k_j}{2m} \right)
      \delta_{g_i g_j},
\end{equation}
where $A_{ij}$ is the weight of the edge connecting nodes $i$ and $j$,
and $A_{ij} = 0$ otherwise; $k_i = \sum_j A_{ij}$ is the weighted degree
of node $i$; $m = \tfrac{1}{2} \sum_{ij} A_{ij}$ is the total edge
weight; $\delta_{g_i g_j}$ is the Kronecker delta, which takes value
$1$ if $g_i = g_j$ and $0$ otherwise; and $g_i$ is the community
assigned to node $i$.}

{The partition is the one that maximises $Q$, found with the Louvain
algorithm~\cite{blondel2008} as implemented in
\textsc{NetworkX} ~\cite{Hagberg2008}. Louvain is a heuristic that optimises $Q$
through iterative local node moves; a fixed random seed is used to
ensure reproducibility.}

To quantify nestedness, we use the NODF (Nestedness based on Overlap and Decreasing Fill)~\cite{AlmeidaNeto2008, 131766}. By doing so, we quantify the extent to which the interactions of a node form subsets of the interactions of the higher-degree ones. {Specifically, for undirected, unweighted networks NODF is calculated as follows:}

\begin{equation}
\text{NODF}_{uni} = \frac{2}{n(n-1)} \sum_{\substack{i<j \\ k_i \neq k_j}}^n \frac{O_{ij}}{\min(k_i, k_j)}
\label{eq:nodf}
\end{equation}

The quantity $ O_{i,j} = \sum_{k=1}^n A_{ik}A_{jk}$ measures the number of common neighbours between each pair of nodes $i$ and $j$ using the overlap matrix $O_{i,j}$.

{In this way, we obtain the average, over all pairs of nodes, of the overlap normalised by the lesser degree of the pair. This normalisation captures hierarchy, because it tests whether the neighbours of the less-connected node form a subset of the more-connected ones.} For example, if all of the smaller node’s neighbours are also linked to the larger one, the ratio equals 1 (perfect nestedness). While if only some are shared, it lies between 0 and 1. Then, the average of this measure across all node pairs quantifies how much the network approaches the ideal nested (hierarchical) structure, where low-degree nodes are entirely embedded into the neighbourhoods of high-degree ones. Prior to computing nestedness, the weighted bipartite and co-occurrence  networks were binarised by setting all non-zero edge weights to one. 
{Eq.~(\ref{eq:nodf}) was applied to the full $n \times n$ binary adjacency matrix of both network types, with $n = C$ for the unipartite co-occurrence network and $n = R + C$ for the bipartite user--hashtag network, where $R$ and $C$ denote the numbers of users and hashtags, respectively.}

\subsection{{$\epsilon^2$ collapse-quality metric}}
\label{sec:epsilon}

{To evaluate the collapse between the original clustering spectrum and those obtained after the procedure of degree-thresholding renormalisation, we computed the distance between them by means of the $\epsilon^2_{cco}$ collapse-quality metric. Following the definition introduced in~\cite{aliakbarisani2025clustering}, $\epsilon^2_{cco}(k)$ is the average of the squared relative differences between each clustering spectrum curve in the DTR flow and that of the original network, computed using exponential binning with $n_{bins}$ bins within the shared range of the $x$-axis for the two curves. Specifically:}

{
\begin{equation*}
    \epsilon^2_{cco} = \frac{1}{n+1}\cdot\frac{1}{n_\text{bins}}\sum_{k_T}\sum_{i=1}^{n_\text{bins}}\Bigg(\frac{\text{bin value}_i^{org}-\text{bin value}_i^{k_T}}{\text{bin value}_i^{org}}\Bigg)^2.
\end{equation*}
}
Here, we used $n_\text{bins} = 20$.  {Smaller $\epsilon^2_{cco}$ values indicate a good collapse of the rescaled curves, i.e.\ self-similar behaviour.} 
\subsection{Geometric model and network embedding}

To investigate further whether the observed self-similarity at CTWs corresponds to an underlying geometric structure, we employed D-Mercator \cite{Jankowski2023}, a model-based embedding method that maps real networks into the $(D+1)$-dimensional hyperbolic space based on the $\mathbb{S}^D$ model~\cite{Serrano2008}. D-Mercator implements a maximum-likelihood inference procedure that assigns each node {$1+D$} hidden coordinates: a popularity coordinate corresponding to a radial position in the latent hyperbolic space that correlates with node degree, and similarity coordinates corresponding to angular positions in a D-dimensional sphere that determine community structure. For simplicity, we used the two-dimensional version, where every node is characterised by a popularity variable or hidden degree $\kappa_i$ and an angular position $\theta_i$ in the one-dimensional sphere representing the similarity space. In $D=1$, D-Mercator infers the popularity-similarity coordinates of the $\mathbb{S}^1$ model~\cite{Serrano2008} or, equivalently, the $\mathbb{H}^2$ formulation~\cite{krioukov2010hyperbolic} in the hyperbolic plane, and has been shown to accurately reproduce complex network topologies whilst remaining computationally tractable. The method assumes that the connection probability between nodes $i$ and $j$ follows:

\begin{equation}
p_{ij}\!=\!\frac{1}{1+\left( \frac{R\Delta\theta_{ij}}{\mu\kappa_i \kappa_j}\right)^{\beta}},
\end{equation}
where $R$ is the radius of the similarity circle, $\Delta \theta_{ij}$ is the angular distance between the nodes, $\mu$ controls the average degree, and $\beta$ controls the level of clustering quantifying the coupling between the network topology and the underlying geometry. 

Prior to embedding, we filtered each CTW network to remove hashtags tweeted by fewer than $N$ users, as low-activity nodes introduce noise that degrades inference quality. The threshold $N$ was selected based on network size and density: after testing $N \in \{3, 5, 7, 10\}$, we determined that $N=3$ for \textit{No al tarifazo} and \textit{Charlie Hebdo} and $N=5$ for \textit{9n} provides the optimal trade-off between noise reduction and preservation of structural features (see Table SI.1 for filtering statistics).

The quality of each embedding was assessed by comparing the empirical connection probability $P(\chi)$ as a function of rescaled hyperbolic distance with the theoretical prediction from the inferred model. High-quality embeddings exhibit smooth decay of $P(\chi)$ with distance, indicating that the network structure is well-explained by the latent hyperbolic geometry.

\vspace{1cm}
\section*{Acknowledgements}
M.A.S. acknowledges support from Grant No. PID2022-137505NB-C22 funded by MCIN/AEI/10.13039/501100011033 and by ERDF/EU and Grant No. PID2025-169261NB-C21 funded by MICIU/AEI/10.13039/501100011033. {YG acknowledges the grant obtained from Labex MME-DII (Grant No. ANR reference 11-LABX-0023-01.})

\bibliography{bibliography}
\end{document}